\documentclass[9pt,twocolumn,twoside]{opticajnl}
\journal{opticajournal} 

\setboolean{shortarticle}{true}
\usepackage{graphicx} 

\title{Fabrication-Aware Inverse Design With Shape Optimization For Photonic Integrated Circuits}

\author[1]{Shaheer Khan*}
\author[1]{Mustafa Hammood}
\author[1]{Nicolas A. F. Jaeger}
\author[1]{Lukas Chrostowski}

\affil[1]{Department of Electrical and Computer Engineering, University of British Columbia, 2332 Main Mall, Vancouver, BC, V6T1Z4, Canada}

\affil[*]{shaheer@ece.ubc.ca, lukasc@ece.ubc.ca}

\begin{abstract}
Inverse design (ID) is a computational method that systematically explores a design space to find optimal device geometries based on specific performance criteria. In silicon photonics, ID often leads to devices with design features that degrade significantly due to the fabrication process, limiting the applicability of these devices in scalable silicon photonic fabrication. We demonstrate a solution to this performance degradation through fabrication-aware inverse design (FAID), integrating lithography models for deep-ultraviolet (DUV) lithography and electron beam lithography (EBL) into the shape optimization approach of ID. A Y-branch and an SWG-to-strip converter were generated and fabricated with this new approach. Simulated and measured results verify that the FAID yields devices with up to 0.6 dB lower insertion loss per device. The modified workflow enables designers to use ID to generate devices that adjust for process bias predicted by lithography models. 
\end{abstract}

\setboolean{displaycopyright}{false} 

\begin{document}

\maketitle

As silicon photonics matures, device performance is increasingly being pushed toward the limits of current technology~\cite{shekhar2024roadmapping}. Traditional methods have long been the foundation of photonic device design, using well-established principles and designer expertise to produce reliable devices. However, the reliance on intuition and the inefficient use of computational power make them unsuitable for exploring the vast design parameter space necessary to further optimize performance. To address these challenges, inverse design (ID) methods have garnered growing interest within the photonics community~\cite{molesky2018inverse}. ID typically employs gradient-based optimization to efficiently navigate the parameter space, often yielding devices that can potentially outperform traditionally designed counterparts~\cite{pan2023deep,hughes2018adjoint}.

Despite their potential, ID devices typically suffer from greater fabrication-induced performance degradation than traditionally designed devices. Photonic device fabrication is subject to deterministic and/or probabilistic process variations that can yield a fabricated device that differs from the ideal geometry and leads to performance degradation \cite{wang2012lithography,xing2022capturing,mistry2019effect}. These process variations include corner rounding, proximity effects, line shortening, and sidewall roughness \cite{mack2000corner, xing2018accurate}, among others. Such process variations are particularly problematic for devices with relatively small feature sizes, such as subwavelength gratings (SWGs)~\cite{yun2019broadband}, devices based on Bragg gratings, such as contra-directional couplers~\cite{lin2019computational}, and topologically optimized devices~\cite{hammond2022high}. Performance degradations resulting from such process variations are especially pronounced in deep-ultraviolet (DUV) lithography, posing challenges to the large-scale manufacturing of devices with subwavelength feature sizes.

For instance, in \cite{wang2012lithography}, integrated Bragg gratings fabricated using 193 nm DUV lithography experienced significant lithographic-induced performance degradation. The designed corrugations, intended to be square, trapezoidal, or triangular, with widths of 10-40 nm, are severely rounded during fabrication. As a result, the experimentally measured bandwidths are narrower than the design values.

In traditional photonic design, designers can anticipate and mitigate performance degradation caused by fabrication variations by relying on their understanding of these processes and employing lithography models. These models predict how deterministic lithographic process variations will alter the as-designed geometry~\cite{lin2019computational,gostimirovic2022deep}. The resulting output is referred to as the 'lithography-predicted geometry', while the as-designed geometry is referred to as the 'ideal geometry.' Current ID optimizers typically optimize a figure-of-merit (FOM) based on the ideal geometry, and consequently do not capture the fabrication-induced performance degradation, highlighting the need for optimization methods that incorporate these considerations.


In this paper, we introduce fabrication-aware inverse design (FAID), a novel method that incorporates lithography models into inverse design. By incorporating predictions of the fabrication-induced process variations, FAID identifies design parameters that adjust for these process variations and mitigate performance degradation. Through simulation and measurement results, we validate an application of FAID with shape optimization~\cite{lalau2013adjoint}, demonstrating substantial improvements over standard ID methods that do not incorporate fabrication-aware techniques.


The design parameter space in photonic ID often contains many solutions that lead to designs that are challenging to fabricate \cite{hammond2022high}. Extensive research has been devoted to constraining the parameter space, largely through the use of filters, constraints, and penalties \cite{hammond2022high,zhou2015minimum,michaels2018leveraging}. Solutions within this constrained or transformed parameter space will pass foundry specific design rule checks. However, since these optimizers still calculate the FOM from an ideal geometry, the solutions do not directly address fabrication-induced performance degradation. Designers may mitigate this issue by tuning constraints to ensure designs with large and smooth feature sizes: however, this can overly constrain the parameter space and result in sub-optimal designs.

\begin{figure}[h]
\centering
\includegraphics[width=3.0in]{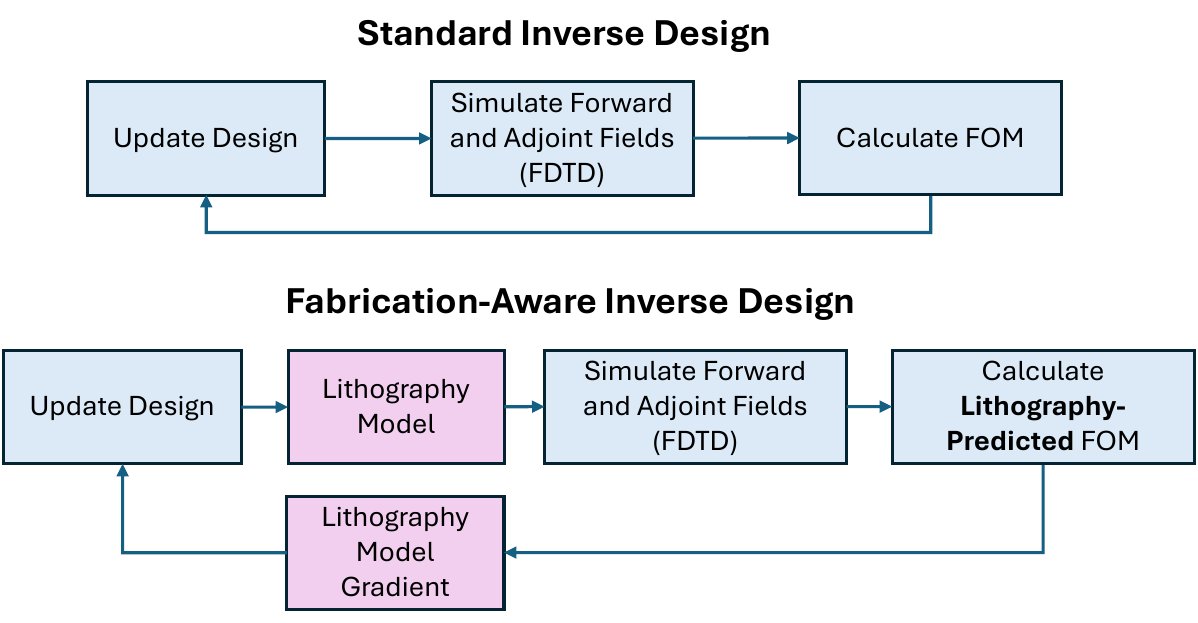}
\caption{Block diagram of the lithography integration with numerical method of calcuating shape gradients}
\label{fig:blockdiagramid}
\end{figure}

FAID leverages lithography models to calculate the FOM based on the lithography-predicted geometry. Rather than imposing constraints on the design parameters, this method transforms the parameter space to be 'fabrication-aware'. The optimal parameters identified in this transformed parameter space will mitigate performance degradation caused by the deterministic process variations captured by the lithography model. A high-level depiction of the workflow is shown in Fig. \ref{fig:blockdiagramid}.

Some lithography models support optical proximity correction (OPC), which adjusts the mask design to compensate for predicted process variations. However, these models are largely proprietary, designed for microelectronics, and often inaccessible to designers since they are typically applied at the mask level. Additionally, since OPC does not account for device performance, the corrected geometry may be non-functional after fabrication due to the omission of critical design features\cite{zhou2014topology}. FAID overcomes these challenges by optimizing directly for performance.

Since ID is most effective when using gradient-based optimization with the adjoint method \cite{lalau2013adjoint}, careful handling of the gradients is crucial to integrate FAID correctly. The adjoint method for photonic ID is described by \cite{pan2023deep}:

\begin{equation}
\begin{aligned}
\frac{dF}{dp}=\underbrace{(-\frac{\partial F}{\partial x}A^{-1})}_{\text{adjoint solution}}\frac{\partial A}{\partial p}x
\end{aligned} \label{eq:fulladjoint}
\end{equation}

\noindent where the objective function $F$ depends solely on a state variable $x$, which in turn is dependent on parameters $p$. $x$ is calculated from the forward simulation, while $(-\frac{\partial F}{\partial x}A^{-1})$ is calculated from the adjoint simulation and utilizes the solution of the forward simulation as a source. The product of the forward and adjoint solutions is referred to as the sensitivity field.

The lithography model can be described as a transformation $g$ on the base parameters $p$. For a differentiable model, this integration is, by the chain rule:

\begin{equation}
\begin{aligned}
\frac{dF}{dp}=\frac{\partial F}{\partial x}\frac{\partial x}{\partial g}\frac{\partial g}{\partial p}
\end{aligned} \label{chain_rule_Litho}
\end{equation}

\noindent indicating that the sensitivity field should be calculated with the adjoint method using the lithography-predicted parameters, the result of which is then multiplied by the Jacobian of the lithography model transformation. 

However, many lithography models are not differentiable. These models can still be used if the gradients are calculated by a finite approximation method, perturbing each parameter and evaluating the sensitivity field~\cite{johnson2002perturbation}.


Two different models were integrated into our proposed FAID method to demonstrate the concept. For DUV lithography, the computational lithography model developed at the University of British Columbia \cite{lin2019computational}, using Mentor Graphics Calibre, was used. For electron beam lithography (EBL), we employed PreFab, a differentiable neural network based model developed at the National Research Council Canada and McGill University\cite{gostimirovic2022deep}. 

\begin{figure}[h]
\centering
\includegraphics[width=3.2in]{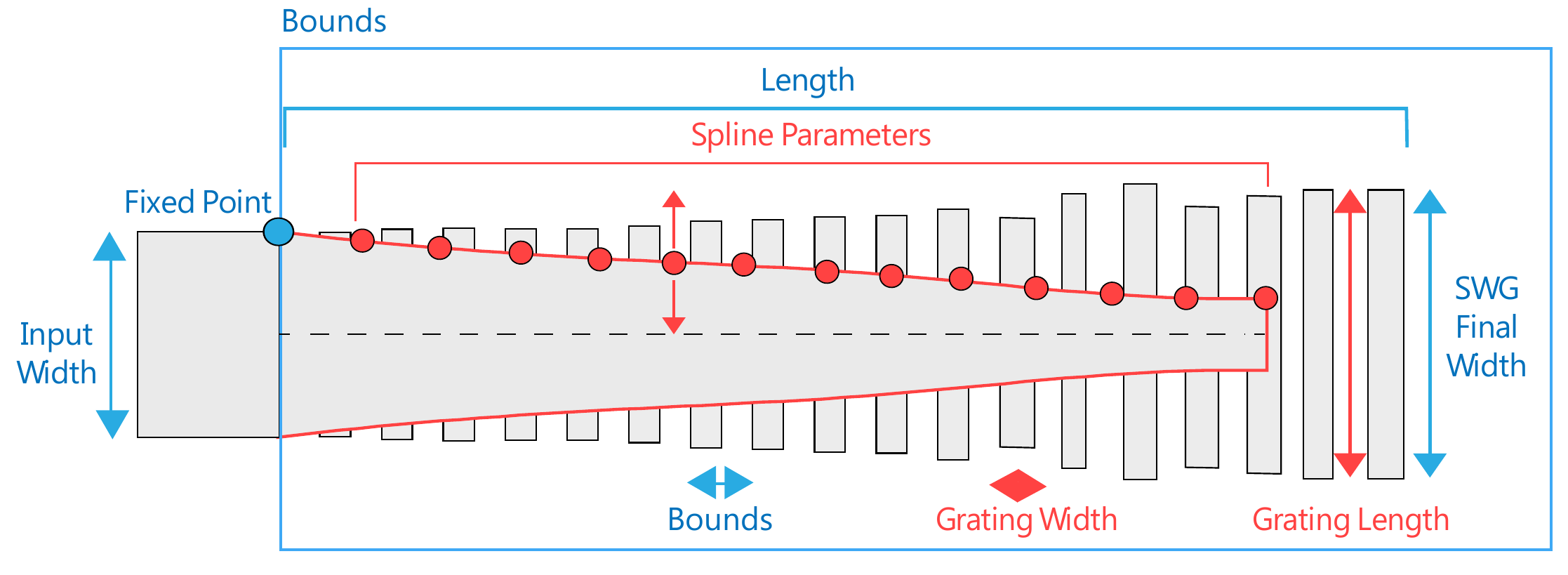}
\caption{Parameterization of the SWG-to-strip converter.}
\label{fig:SWGPara}
\end{figure}

Two device types were used to assess the validity of FAID. First, a Y-branch was chosen because of its widespread use in inverse design research \cite{zhang2013compact,lalau2013adjoint}. Although applying smoothness penalties reduces the likelihood of significant performance degradation \cite{michaels2018leveraging}, we demonstrate that FAID still offers advantages. Second, a taper that converts from a sub-wavelength grating (SWG) waveguide\cite{cheben2006subwavelength} to a strip waveguide (SWG-to-strip converter) was selected due to its dependence on minimum attainable feature sizes, its importance in several applications \cite{donzella2014sub}, and its large parameter space. We allow the width and length of individual gratings and the boundary of the waveguide bridge to vary. The parameterization of the SWG-to-strip converter is shown in Fig. \ref{fig:SWGPara}. 


\vspace{-1pt} 

The validation methodology for FAID involves the following steps:

\begin{enumerate}{}{}
\item{Two devices are generated with shape optimization, one with standard ID and one with FAID. }
\vspace{-1pt}
\item{Both devices are passed to the lithography predictor model, and then simulated to compare the lithography-predicted geometry’s performance.}
\vspace{-1pt}
\item{Both devices are fabricated, and their measurements are compared.}
\end{enumerate}

\begin{figure}[h!]
\centering
\includegraphics[width=3in]{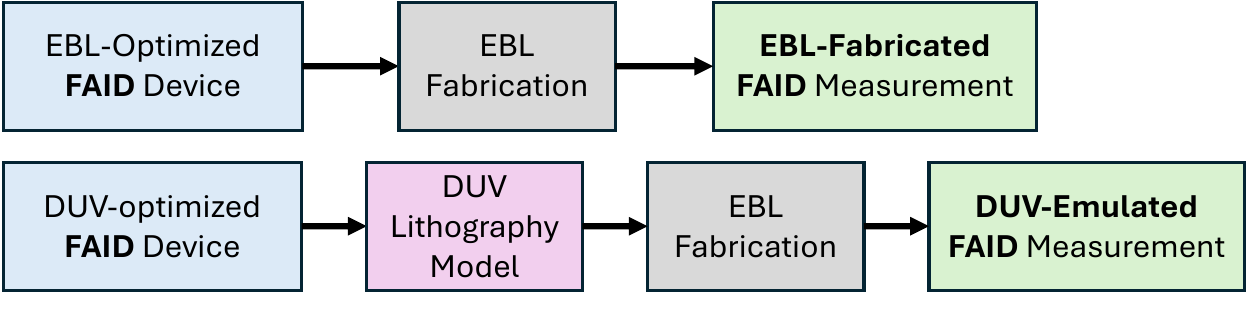}
\caption{Fabrication validation process for the FAID device. This is repeated for the standard ID device to compare measurement results.}
\label{fig:fabfad}
\end{figure}

This procedure is applied to both the Y-branch and SWG-to-strip converter, and repeated for both the EBL and DUV lithography processes. A simulation test is deemed successful if the FAID device's lithography-predicted performance surpasses that of the standard ID device. The experiment is then validated by comparing the post-fabrication measurements of the FAID device and the ID device. The process for obtaining measured data is shown in Fig. \ref{fig:fabfad}. To test the device performance with DUV fabrication, the DUV-lithography model is applied to the ID and FAID devices prior to EBL fabrication in order to emulate the DUV process \cite{hammood2023emulation}. 

\begin{figure}[h]
\centering
\includegraphics[width=2.7in]{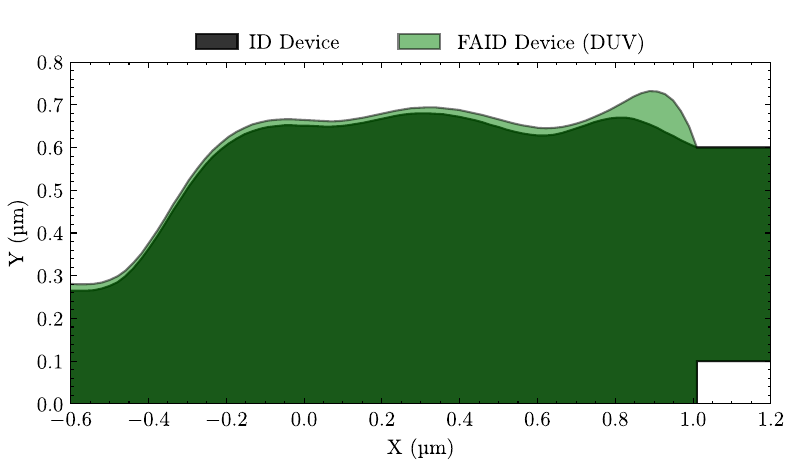}
\caption{Illustration of the top halves of symmetrical FAID and ID Y-branches. The FAID device is optimized for DUV. Note that only the boundary is optimized, hence there are no changes between the output waveguides.}
\label{fig:ylitho}
\end{figure}

Fig. \ref{fig:ylitho} contrasts an ID optimized Y-branch with a FAID optimized Y-branch that leverages the DUV model. It can be seen that the lithography optimization compensates for the under-etch predicted by the lithography model, as well as an the adjustment at the edge of the Y-branch to ensure smoother transitions to the output waveguides. 

\vspace{-1pt}

\vspace{-1pt}

\begin{figure}[h]
\centering
\includegraphics[width=3.1in]{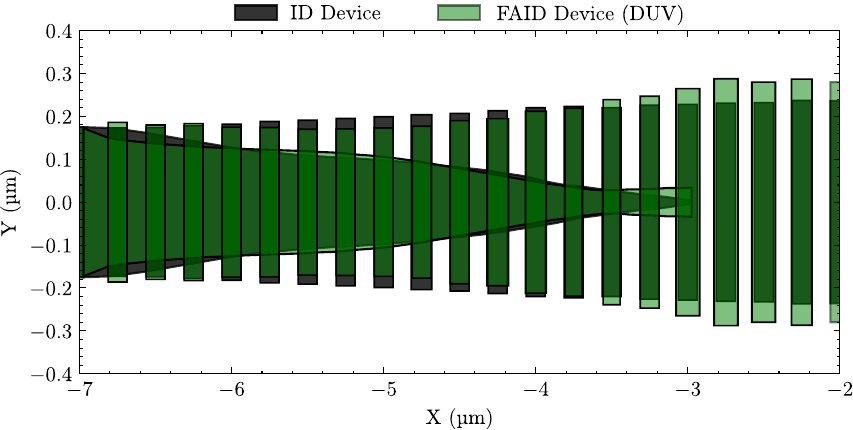}
\caption{Illustration of the FAID and ID SWG-to-strip converters. The FAID device is optimized for DUV lithography.}
\label{fig:SWGlitho}
\end{figure}

Similar intuitive adjustments can be observed for the SWG-to-strip device illustrated in Fig. \ref{fig:SWGlitho}. Since the lithography model is utilized throughout the optimization process, it is very likely that the FAID device will reach a local optimum with an altogether different set of parameters. Many of these parameters contribute to reduced performance degradation, but it should be noted that due to the nonconvexity of the parameter space, some parameters may not be contributing to improved lithography-predicted performance.

\begin{figure}[h!]
\centering
\includegraphics[width=2.6in]{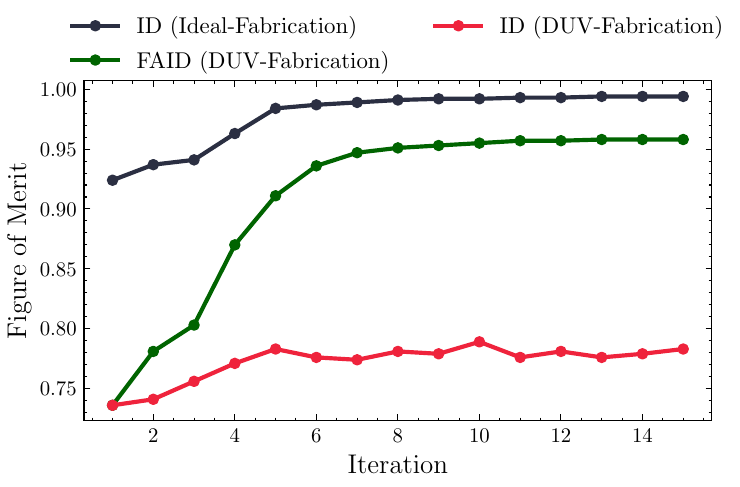}
\caption{FOM at each iteration of the optimization process of the SWG-to-strip converter.}
\label{fig:lbfgs_opt}
\end{figure}

Fig. \ref{fig:lbfgs_opt} illustrates the optimization process for the SWG-to-strip converter with both ID and FAID. The dark blue line represents the ID device's ideal (i.e.: without lithography effects) performance at each iteration of optimization. At every step of the optimizer, we also plot, in red, the performance of the DUV-predicted ID device. After the first few iterations, further improvements in the FOM calculated from the ideal geometry do not yield improvements in the FOM calculated from the DUV-predicted geometry, ultimately yielding poor results. This reflects a situation where standard ID will not yield a design which works well once fabricated, even though in simulation the ideal performance is excellent. In contrast, the performance of the \textit{DUV-predicted} FAID device, indicated by the green line, yields consistent improvement at each iteration, ultimately approaching the 
performance of the \textit{ideal} ID geometry.

\vspace{-1pt}

\begin{table}[h!]
\centering
\caption{\bf Simulated Insertion Loss(dB) at 1320 nm Comparison: FAID vs. Standard ID}
\resizebox{0.99\linewidth}{!}{%
\begin{tabular}{lccc}
\hline
\textbf{Device} & \textbf{ID (Ideal)} & \textbf{ID (Predicted)}& \textbf{FAID (Predicted)} \\
\hline
\textbf{Y-branch (EBL)} & 0.067 & 0.072 & 0.067 \\
\textbf{Y-branch (DUV)} & 0.067 & 0.089 & 0.068 \\
\textbf{SWG-to-strip (EBL)} & 0.034 & 0.072 & 0.036 \\
\textbf{SWG-to-strip (DUV)} & 0.034 & 0.836 & 0.111 \\
\hline
\end{tabular}%
}
\end{table}

The simulation results are presented in Table 1. The 'ideal ID' results reflect the simulated performance of ID assuming an ideal fabrication process. Once the process variations predicted by the lithography model are factored in, the simulated performance degrades considerably, especially in devices with smaller feature sizes. In contrast, the performance of the lithography-predicted FAID devices show a substantial mitigation of the fabrication induced performance degradation. This mitigation is especially significant for the DUV-predicted FAID SWG-to-strip converter, yielding a 0.7 dB improvement compared to the DUV-predicted ID device.

\begin{figure}[h!]
\centering
\includegraphics[width=2.6in]{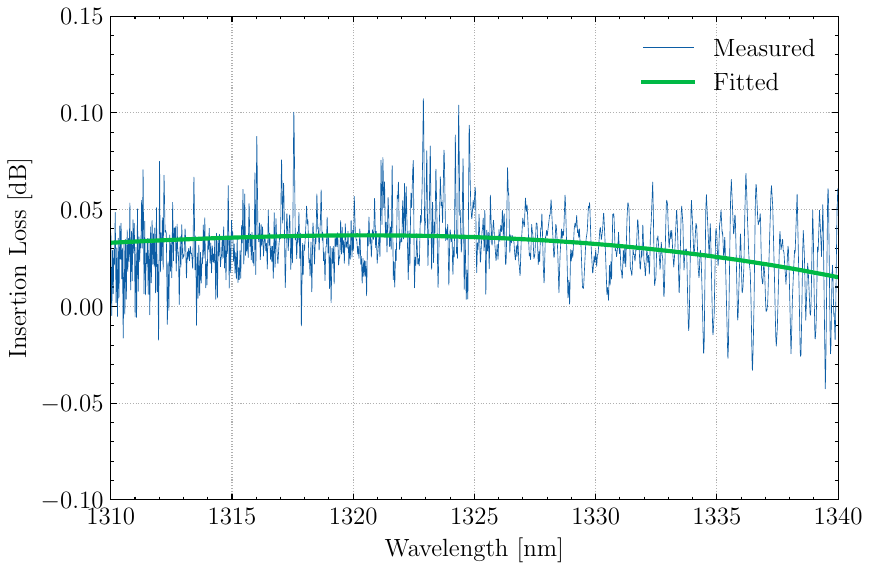}
\caption{Insertion loss extraction of the EBL-optimized SWG-to-strip converters}
\label{fig:cutback_measurement}
\end{figure}

The EBL foundry used to validate our work was Applied Nanotools, which uses a JEOL JBX-8100FS EBL system\cite{chrostowski2019silicon}.

The measured insertion loss extracted using the cutback method for the EBL-optimized SWG-to-strip is depicted in Fig. \ref{fig:cutback_measurement}. This device was measured to have an insertion loss of 0.033 dB compared to 0.081 dB for the ID device. 

\begin{figure}[h!]
\centering
\includegraphics[width=2.8in]{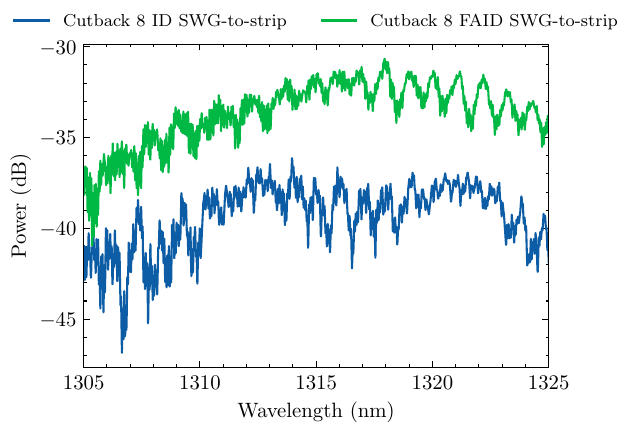}
\caption{Transmission spectra showing the measured loss of 8 DUV-emulated FAID SWG-to-strip converters compared to the same number of ID devices. The setup consists of an input and output grating coupler, strip waveguides, 4 pairs of converters, and a detector.}
\label{fig:DUVunoptopt}
\end{figure}

\vspace{-2pt}

The transmission spectra for both the ID and FAID-with-DUV SWG-to-strip converters are presented in Fig. \ref{fig:DUVunoptopt}. It can be seen that the FAID devices optimized for DUV devices suffer far less performance degradation.

\begin{table}[h!]
\centering
\caption{\bf Average Measured Insertion Loss (dB) from 1305-1325 nm: FAID vs. ID}
\resizebox{0.75\linewidth}{!}{%
\label{tab1}
\begin{tabular}{lcc}
\hline
\textbf{Device} & \textbf{ID} & \textbf{FAID} \\
\hline
\textbf{Y-branch (EBL) } & 0.076 & 0.077 \\
\textbf{Y-branch (DUV Emulated)} & 0.100 & 0.080 \\
\textbf{SWG-to-strip (EBL)} & 0.081 & 0.033 \\
\textbf{SWG-to-strip (DUV Emulated)} & 1.321 & 0.611 \\
\hline
\end{tabular}%
}
\end{table}

Table 2 summarizes the measurement results, as obtained using the cutback method. In all cases, the FAID devices match or substantially exceed the performance of the ID devices. In particular, the FAID SWG-to-strip converter optimized for DUV lithography achieves a 0.61 dB reduction in insertion loss compared to its ID counterpart. The EBL-optimized FAID SWG-to-strip converter shows exceptionally low loss compared to previous designs~\cite{donzella2014sub}. Even for devices with relatively smooth features like the Y-branch, FAID considerably reduces expected performance degradation with DUV lithography. 

In conclusion, we presented a method to address fabrication-induced performance degradation in inverse design by integrating lithography models into the ID process. We demonstrate that standard ID can result in devices with poor real-world performance due to deviations from the ideal geometry as a result of fabrication. Through both simulation and measurement results of devices generated with shape optimization, we validate that FAID effectively mitigates fabrication-induced performance degradation, particularly for DUV-lithography. This represents a significant step towards leveraging ID for scalable, commercial silicon photonic applications.

Future work should extend this method to topology optimization. The extremely small feature sizes of these devices are likely to cause significant deterioration in the FOM as a result of DUV fabrication, which the FAID technique can resolve. Co-optimizations using the FAID technique with uncertainty predictions from lithography models can yield devices that are also robust to probabilistic process variations.

\begin{backmatter}
\bmsection{Funding} This work was supported by the Natural Sciences and Engineering Research Council of Canada (NSERC).

\bmsection{Acknowledgments} Most measurements were conducted by Omid Esmaeeli at the University of British Columbia.

\bmsection{Disclosures} The authors declare no conflicts of interest.

\bmsection{Data availability} Data underlying the results presented in this paper are not publicly available at this time but may be obtained from the authors upon reasonable request.

\end{backmatter}

\bibliography{main}

\bibliographyfullrefs{main}


\end{document}